\documentclass[aps,prl,twocolumn,superscriptaddress]{revtex4}

\usepackage[dvips]{graphicx}
\usepackage{amsmath}
\usepackage{url}
\usepackage{bm}
\usepackage{color}
\usepackage{comment}

\usepackage[dvipdfmx]{hyperref}
\hypersetup{
    colorlinks=true,
    citecolor=blue,
    linkcolor=blue,
    urlcolor=blue,
}
\newcommand{\red}[1]{\textcolor{magenta}{#1}}

\begin{document}
\title{
   Switchable Polarization in an A-site Deficient Perovskite through Vacancy and Cation Engineering
}
\author{Suguru Yoshida}
\email[e-mail:]{suguru.yoshida0224@gmail.com}
\affiliation{Department of Energy and Hydrocarbon Chemistry, Kyoto University, Kyoto 615-8510, Japan}
\author{Olivier Hernandez}
\affiliation{Institut des Mat\'{e}riaux de Nantes Jean Rouxel,  Nantes Universit\'{e}, Nantes F-44000, France}

\author{Jinsuke Miyake}
\affiliation{Department of Material Chemistry, Kyoto University, Kyoto 615-8510, Japan}
\author{Kei Nakayama}
\affiliation{Institute of Engineering Innovation, School of Engineering, The University of Tokyo, Bunkyo, Tokyo 113-8656, Japan}
\affiliation{Nanostructures Research Laboratory, Japan Fine Ceramics Center, Nagoya, Aichi 456-8587, Japan}
\author{Ryo Ishikawa}
\affiliation{Institute of Engineering Innovation, School of Engineering, The University of Tokyo, Bunkyo, Tokyo 113-8656, Japan}
\author{Hajime Hojo}
\affiliation{Department of Advanced Materials Science and Engineering, Kyushu University, Fukuoka 816-8580, Japan}
\author{Yuichi Ikuhara}
\affiliation{Institute of Engineering Innovation, School of Engineering, The University of Tokyo, Bunkyo, Tokyo 113-8656, Japan}
\affiliation{Nanostructures Research Laboratory, Japan Fine Ceramics Center, Nagoya, Aichi 456-8587, Japan}
\author{Venkatraman Gopalan}
\affiliation{Materials Research Institute and Department of Materials Science and Engineering, Pennsylvania State University, University Park, PA 16802, USA}
\author{Katsuhisa Tanaka}
\author{Koji Fujita}
\affiliation{Department of Material Chemistry, Kyoto University, Kyoto 615-8510, Japan}
%

\begin{abstract}
   While defects are unavoidable in crystals and often detrimental to material performance, they can be a key ingredient for inducing functionalities when tailored.
   Here, we demonstrate that an A-site-deficient perovskite Y$_{1/3}$TaO$_3$ exhibits room-temperature ferroelectricity in a $Pb2_1m$ phase,
   enabled by ordered vacancies coupled with TaO$_6$ octahedral rotations.
   Defect-ordered perovskites are frequently trapped in centrosymmetric incommensurate states due to competing structural instabilities;
   we circumvent this by favoring rotational over polar instability through compositional selection.
   Unlike canonical improper ferroelectrics that are \textit{ferrielectric}, the vanishing dipoles on vacancy layers in Y$_{1/3}$TaO$_3$ allow for a net ferroelectric alignment of local dipoles,
   resulting in enhanced polarization.
   Upon heating, Y$_{1/3}$TaO$_3$ transforms to a paraelectric incommensurate phase at $\simeq$750 K, whose atomic arrangement mirrors the domain topology observed in hybrid improper ferroelectrics.
   Superspace analysis of the modulated phase reveals a route to improve room-temperature polarization, achieved through epitaxial strain, as confirmed by our lattice-dynamics calculations.
   This defect-ordering strategy should be generalizable to other improper ferroelectrics, including magnetoelectric multiferroics,
   providing a pathway to amplify otherwise limited macroscopic polarization.
\end{abstract}


\maketitle
\section{Introduction}
Schottky defects are entropically inevitable at finite temperatures.
Left uncontrolled, they degrade material performance~\cite{liu2009PRB,qiu2013NC,heremans2017NRMa}.
Yet rationally introduced and organized defects can unlock physical properties inaccessible in perfect crystals.
For instance, oxygen vacancies convert insulating ZrO$_2$ into a fast-ion conductor~\cite{inaba1996SSI,steele2001N}.
In SrTiO$_3$, fine-tuning of Sr-vacancy concentration enhances its thermoelectric figure of merit~\cite{lu2016CM}.
Likewise, in cuprates, carefully engineered oxygen nonstoichiometry has raised the superconducting transition temperature above the boiling point of nitrogen~\cite{wu1987PRL,strobel1987N,lee2006RMP}.
Defects can be created by a range of synthetic approaches~\cite{arandiyan2021CSR}, including  heterovalent substitution~\cite{inaguma1993SSC,smyth2000SSI}, gas atmosphere~\cite{strobel1987N,lee2016SR},
topochemical reaction~\cite{hayward1999JACS,cassidy2019IC}, and chemical potential control~\cite{ohno2018J,borgsmiller2022PE},
and especially when ordered to some extent, these thermodynamic imperfections become powerful tuning knobs in the design of functional solids~\cite{kawakami2009NC,yamamoto2020NC}.

\begin{figure}
   \includegraphics[width=8.6cm]{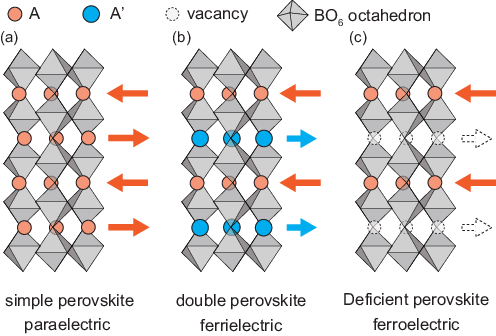}
   \caption{
   Schematic illustration of (a) a simple perovskite, (b) A/A$^{\prime}$ layer-ordered double perovskite,
   and (c) A-site deficient perovskite with an alternate ordering of vacancies,
   all of which exhibit octahedral rotations corresponding to the Glazer notation of $a^-a^-c^+$~\cite{glazer1972ACB}.
   In panels (b) and (c), the asymmetry introduced by the distinct adjacent A layers leads to the absence of inversion centers at the B cations.
   Red and blue arrows show local electric dipoles arising from A and A$^{\prime}$ cation displacements. 
   Dotted arrows in panel (c) indicate vanishing local polarization due to the complete absence of cations.
   }
\label{fig:structures_scheme}
\end{figure}

Because ordered vacancies can lower the crystal symmetry, they are naturally compatible with functionalities, such as ferroelectricity,
relying on the lifting of spatial inversion~\cite{mishra2014NL,young2017JACS,zhang2024NM,yang2025JACS}.
Layer-selective A-site vacancy ordering reproduces the A/A$^{\prime}$ double-perovskite architecture identified by theory
as a prerequisite for hybrid improper ferroelectricity~\cite{benedek2011PRL,mulder2013AFM,young2013CM} (Fig.~\ref{fig:structures_scheme}).
This specific arrangement, by inherently breaking the inversion symmetry at the B sites,
allows two non-polar octahedral rotations to couple trilinearly with a polar displacement.
Upon a particular rotational pattern of BO$_6$ octahedra, the A-site cations shift in an antiparallel fashion.
Those layer dipoles cancel out in conventional ABO$_3$ perovskites (Fig.~\ref{fig:structures_scheme}a) but yield a net polarization,
leading to \textit{ferrielectricity} in double perovskites~\cite{benedek2012JSSC,mulder2013AFM} (Fig.~\ref{fig:structures_scheme}b) and ferroelectricity in defect-ordered perovskites (Fig.~\ref{fig:structures_scheme}c). 
Indeed, A-site deficient perovskites A$_{1 - x}$BO$_3$~\cite{keller1965JoIaNC,rooksby1965JACS,labeauu1982ACSA,ebisu2000JoPaCoS,zhou2008CM,azough2010JACS} such as Ln$_{1/3}$NbO$_3$ and Ln$_{2/3}$TiO$_3$ (Ln: lanthanides) have been proposed
as candidate platforms to explore polar phenomena~\cite{beqiri2019CC}.
However, these materials often settle into compromised centrosymmetric and incommensurate phases owing to a competition between rotational instability and B-cation off-centering~\cite{abakumov2013CM,beqiri2019CC,labeauu1982ACSA}.
Consequently, conclusive experimental evidence of a polar structure, let alone switchable ferroelectricity, has yet to be demonstrated in these systems.

To tip the balance toward a rotation-driven hybrid improper ferroelectric state, we seek a design strategy that simultaneously
(i) suppresses the off-centering displacement
and (ii) stabilizes the rotated structure, all while preserving the vacancy-ordered architecture.
Substituting Ta$^{5+}$ for Nb$^{5+}$ fulfills these criteria:
Ta$^{5+}$ with 5d orbitals is less prone to off-centering distortion,
as exemplified by the contrast between KNbO$_3$ (a displaced polar perovskite) and KTaO$_3$ (an incipient quantum paraelectric perovskite)~\cite{shirane1954PR,rowley2014NP,postnikov1994PRB},
while the larger radial extent of the 5d orbitals is expected to maintain the rotational instability according to the electronic mechanism of octahedral rotations~\cite{yoshida2021PRL,koiso2024PRB}. 
Of these, a compound with A = Y$^{3+}$ is an ideal system to assess the acentricity and ferroelectric nature by optical second harmonic generation (SHG) and hysteresis measurements.
This is due to the absence of f electrons, which ensures optical transparency and may contribute to reduced leakage current.

Guided by this rationale, we investigate Y$_{1/3}$TaO$_3$,
an A-site-deficient perovskite oxide previously reported but lacking definitive structural and property characterization.
Combined analysis by diffraction and optical SHG confirms the defect-ordered perovskite adopts a polar structure,
and hysteresis measurements demonstrate switchable polarization.
This establishes Y$_{1/3}$TaO$_3$ as the first defect perovskite with robust improper ferroelectricity.
We show that upon heating, Y$_{1/3}$TaO$_3$ undergoes a phase transition to an incommensurately modulated paraelectric structure;
the detailed structural characterization of this incommensurate phase provides a strategy to achieve even larger polarization values.
These findings corroborate the utility of defect ordering to activate improper ferroelectricity,
and we anticipate that this ferroelectric---rather than \textit{ferrielectric}---ground state, enabled by defect order,
underscores the potential of vacancy engineering as a general strategy to amplify polarization in improper ferroelectrics, including magnetoelectric multiferroics~\cite{vanaken2004NM}.

\section{Methods}
\subsection{Experimental}
Polycrystalline samples of Y$_{1/3}$TaO$_3$ were synthesized by solid-state reaction.
Reagent-grade Y$_2$O$_3$ (99.99\%) and Ta$_2$O$_5$ (99.9\%) were used as raw materials.
Before weighing, Y$_2$O$_3$ powders were heated at 900~$^{\circ}$C for 12~h to eliminate water and carbon dioxide absorbed in the powders.
Y$_2$O$_3$ and Ta$_2$O$_5$ powders were mixed so as to obtain a Y:Ta cation ratio of 1:3, grounded in an agate mortar, and pressed into a pellet.
The pellet was calcinated at 850~$^{\circ}$C for 12~h. The resultant pellet was ground, thoroughly mixed, pelletized again, and sintered at 1600~$^{\circ}$C for 96~h.

Variable-temperature synchrotron x-ray diffraction (SXRD) data were collected for Y$_{1/3}$TaO$_3$ in the temperature range of 300--1100 K at SPring-8 BL02B2, Japan, using a large Debye--Sherrer camera.
The incident x-ray was monochromated at $\lambda$ = 0.420272 or 0.501146~\AA.
The powder sample was housed in a silica capillary tube with an inner diameter of 0.2~mm and rotated continuously during the measurements to diminish preferred orientation.
High-resolution neutron diffraction (ND) patterns were recorded at different temperatures using $2\theta$-dispersive diffractometer D2B at Institute Laue Langevin, France,
and time-of-flight diffractometer HRPD at the ISIS facility, UK.
For the measurements at D2B, approximately 4.5~g of the sample was housed in an evacuated vanadium can with an inner diameter of 9~mm.
Using incident neutron wavelength of $\lambda$ = 1.594~\AA, diffraction patterns were recorded at 300--1000~K.
Another wavelength of 2.398~\AA\, was also used for collecting the patterns at 300, 800, and 900~K. 
For the measurements at HRPD, approximately 3~g of the sample was put in an air-filled vanadium can with an inner diameter of 6~mm.
Using three detection banks (backscattering bank, 90$^{\circ}$ bank, and low-angle bank),
diffraction patterns were obtained over a time-of-flight range of 30--130~ms corresponding to a $d$-range of 0.65--9.0~\AA.
Structural parameters were refined through Rietveld analysis with \textsc{jana2006}~\cite{petricek2014ZFK-CM} against the SXRD and ND datasets.

Selected-area electron diffraction (SAED) patterns of Y$_{1/3}$TaO$_3$ at 300~K and 900~K were recorded
using JEM-ARM200F and JEM-2010HC transmission electron microscopes (JEOL Ltd.), respectively, operated at 200~kV.
Specimens for the experiments were prepared by dispersing the powdered sample onto holey carbon films supported on Cu grids.
The grid was mounted on a heating holder (JEOL Ltd.) for the high-temperature measurements.
High-angle annular dark-field (HAADF) scanning transmission electron microscopy (STEM) images were obtained at room temperature using JEM-ARM300CF (JEOL Ltd.),
operated at 300~kV.
The probe-forming aperture semiangle was 24~mrad, and the collection semiangle ranged from 90 to 200~mrad.

Optical SHG was measured for the sintered polycrystalline pellet of Y$_{1/3}$TaO$_3$ in reflection geometry using a regeneratively amplified mode-locked Ti:sapphire laser
(800-nm wavelength, 80-fs pulse duration, and 1-kHz repetition rate).
Variable-temperature data were recorded at a heating rate of 10~K/min using a home-built heater.

Electric polarization versus electric field ($P$--$E$) hysteresis loop was measured for a dense polycrystalline pellet (relative density of 97\%) of Y$_{1/3}$TaO$_3$
with a ferroelectric tester (Precision LC, Radiant Technologies) and a high-voltage amplifier (10~kV HVI-SC, Radiant Technologies).
The pellet was prepared from a stoichiometric mixture of the raw powders by a uniaxial press,
cold isostatic pressing, and subsequent sintering at 1600~$^{\circ}$C for 96~h.

Differential scanning calorimetry (DSC) was recorded for the powder sample of Y$_{1/3}$TaO$_3$ from room temperature to 900~K by Rigaku Thermo Plus DSC 8270 at a heating rate of 10~K/min.

\subsection{Computational}
Density functional theory (DFT) calculations were performed with the projector augmented-wave (PAW) method~\cite{blochl1994PRB,kresse1999PRB}
and PBEsol functional~\cite{perdew1996PRL,perdew1997PRL,perdew2008PRL} as implemented in the \textsc{vasp} code~\cite{kresse1993PRB,kresse1993PRBa,kresse1996PRB,kresse1996CMS}.
A plane-wave cutoff energy of 550~eV was used with the standard PAW cutoffs.
To circumvent the challenges originating from the partial distribution of Y$^{3+}$, we assumed that Ca$^{2+}$ fully occupied the alternate A-site layers,
corresponding to the chemical composition of Ca$_{1/2}$BO$_3$ (B = Nb and Ta).
This methodology follows precedents in the literature that modeled the partial occupation in analogous Nb-based perovskites~\cite{beqiri2019CC}.
The following states were treated as valence electrons:
3p and 4s for Ca;
4p, 4d, and 5s for Nb;
5d and 6s for Ta; and
2s and 2p for O.
Under the Monkhorst--Pack scheme~\cite{monkhorst1976PRB}, the mesh sampling in the $k$-space was set to $8 \times 8 \times 4$ and $4 \times 4 \times 4$ for the unit cell of $P4/mmm$ structure with 9 atoms
and its $2 \times 2 \times 1$ supercell, respectively. 
Note that the unit cell of $P4/mmm$ structure is enlarged along the $c$-axis with the metric of $\bm{a}_{\mathrm{pc}} \times \bm{b}_{\mathrm{pc}} \times 2\bm{c}_{\mathrm{pc}}$,
where $\bm{a}_{\mathrm{pc}}$, $\bm{b}_{\mathrm{pc}}$, and $\bm{c}_{\mathrm{pc}}$ represent the primitive lattice vectors of the ideal cubic perovskite.
The lattice vectors and fractional coordinates were optimized until the residual stress and forces were reduced to 0.01~GPa and 1~meV/\AA, respectively. 
Phonon band dispersions were calculated for these supercell structures (2$\bm{a}_{\mathrm{pc}} \times 2\bm{b}_{\mathrm{pc}} \times 2\bm{c}_{\mathrm{pc}}$) by using force constants obtained from the DFT calculations, employing the finite displacement method as implemented in the \textsc{phonopy} code~\cite{togo2015SM}.

\section{Results and Discussion} 
\subsection{Room-temperature polar structure and ferroelectricity}
\begin{figure}
   \includegraphics[width=8.6cm]{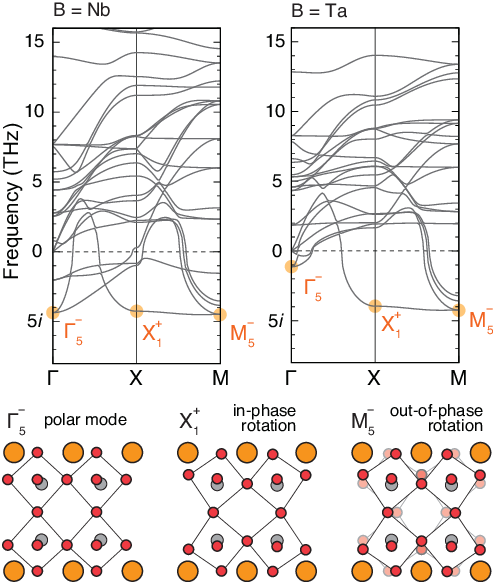}
   \caption{
   Phonon dispersion curves calculated along a high-symmetry path [$\Gamma$(0\,0\,0)--X($\frac{1}{2}$\,0\,0)--M($\frac{1}{2}$\,$\frac{1}{2}$\,0)] for the nondistorted
   $P4/mmm$ structure of Ca$_{1/2}$BO$_3$ (B = Nb and Ta).
   This model represents Y$_{1/3}$BO$_3$ systems, where half of the A-site is randomly occupied by Y$^{3+}$.
   Eigendisplacements for key vibrational modes, including $\Gamma_5^-$, X$_1^+$, and M$_5^-$ modes, are depicted in the lower panel.
   }
\label{fig:phonon}
\end{figure}

To evaluate how the introduction of Ta$^{5+}$ to the B site affects the dynamical property of Y$_{1/3}$BO$_3$,
we calculate and compare phonon dispersion curves for the $P4/mmm$ phases of hypothetical model systems, Ca$_{1/2}$BO$_3$ (B = Nb and Ta; Fig.~\ref{fig:phonon}). 
For the Nb system, a pronounced imaginary frequency ($\simeq$ 4.36$i$~THz) corresponding to the $\Gamma_5^-$ mode indicates a strong propensity toward polar displacement.
In stark contrast, the same $\Gamma_5^-$ mode in Ca$_{1/2}$TaO$_3$ shows a reduced imaginary frequency ($\simeq$ 1.13$i$~THz), reflecting its suppressed polar instability.
However, the imaginary frequencies at the zone-boundary X and M points, corresponding to in-phase and out-of-phase octahedral rotations, respectively,
remain comparable in magnitude for both compounds.
This observation confirms that the stability of rotated structures is well preserved despite the B-site substitution,
enabling the designed unbalancing of polar and rotational instabilities in the Ta-based system.

\begin{figure}
   \includegraphics[width=8.5cm]{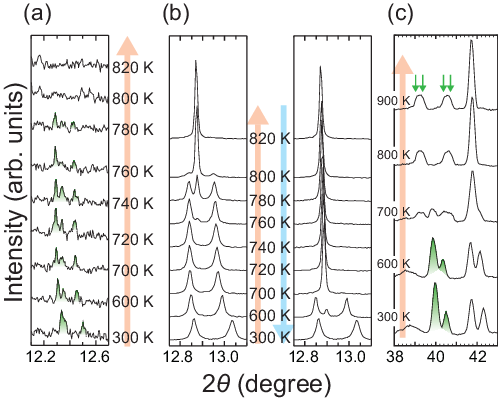}
   \caption{
   Temperature dependence of (a) $\frac{1}{2}$$\frac{1}{2}$3, $\frac{1}{2}$$\frac{3}{2}$1, and $\frac{3}{2}$$\frac{1}{2}$1 superlattice reflections upon heating 
   and (b) 112 reflection upon heating and cooling (SXRD, $\lambda$ = 0.501146~\AA).
   The Miller indices are defined with respect to the $P4/mmm$ unit cell with the possible highest symmetry. 
   (c) Temperature-variable ND patterns ($\lambda$ = 1.594~\AA) highlighting the same superlattice reflections and peak splitting at lower temperatures. 
   The green arrows indicate satellite reflections of the paraelectric phase. 
   }
\label{fig:SLR}
\end{figure}

Prior studies on Y$_{1/3}$TaO$_3$ have reported no atomic structure, with conflicting results regarding its crystal system~\cite{keller1965JoIaNC,rooksby1965JACS,ebisu2000JoPaCoS,kawai2022SM}.
While some of the other Ln$_{1/3}$TaO$_3$ (e.g., Ln = Ho and Er) were assigned to a pyroelectric group $Pmc2_1$ based on SXRD data~\cite{zhou2008CM},
their potential for polarity or ferroelectricity has not been discussed. 
We therefore comprehensively re-examine the room-temperature crystal structure of Y$_{1/3}$TaO$_3$ by ND and SXRD techniques. 
Bragg reflections in the SXRD pattern cannot be captured by the ideal $P4/mmm$ cell with $\bm{a}_{\mathrm{pc}} \times \bm{b}_{\mathrm{pc}} \times 2\bm{c}_{\mathrm{pc}}$ lattice vectors.
Instead, we observed weak $h + \frac{1}{2}$ $k + \frac{1}{2}$ $l$ superlattice reflections
(e.g., $\frac{1}{2}$$\frac{1}{2}$3, $\frac{1}{2}$$\frac{3}{2}$1, and $\frac{3}{2}$$\frac{1}{2}$1 reflections in Fig.~\ref{fig:SLR}a)
and the splitting of \textit{hhl} reflections such as 112 reflection in Fig.~\ref{fig:SLR}b.
Similar features, with the superlattice reflections being more intense, were found in ND patterns (Fig.~\ref{fig:SLR}c),
suggesting that the structural deviation from the $P4/mmm$ structure is likely to relate to lighter elements like oxygen.  
Alternatively, an orthorhombic unit cell with an enlarged cell metric of $\sqrt{2}\bm{a}_{\mathrm{pc}} \times \sqrt{2}\bm{b}_{\mathrm{pc}} \times 2\bm{c}_{\mathrm{pc}}$ successfully indexes all observed SXRD peaks. 
The enlarged cell aligns with recent reports on the Ho and Er analogs, diverging from earlier identifications for Y$_{1/3}$TaO$_3$. 

\begin{figure*}
   \includegraphics[width=17cm]{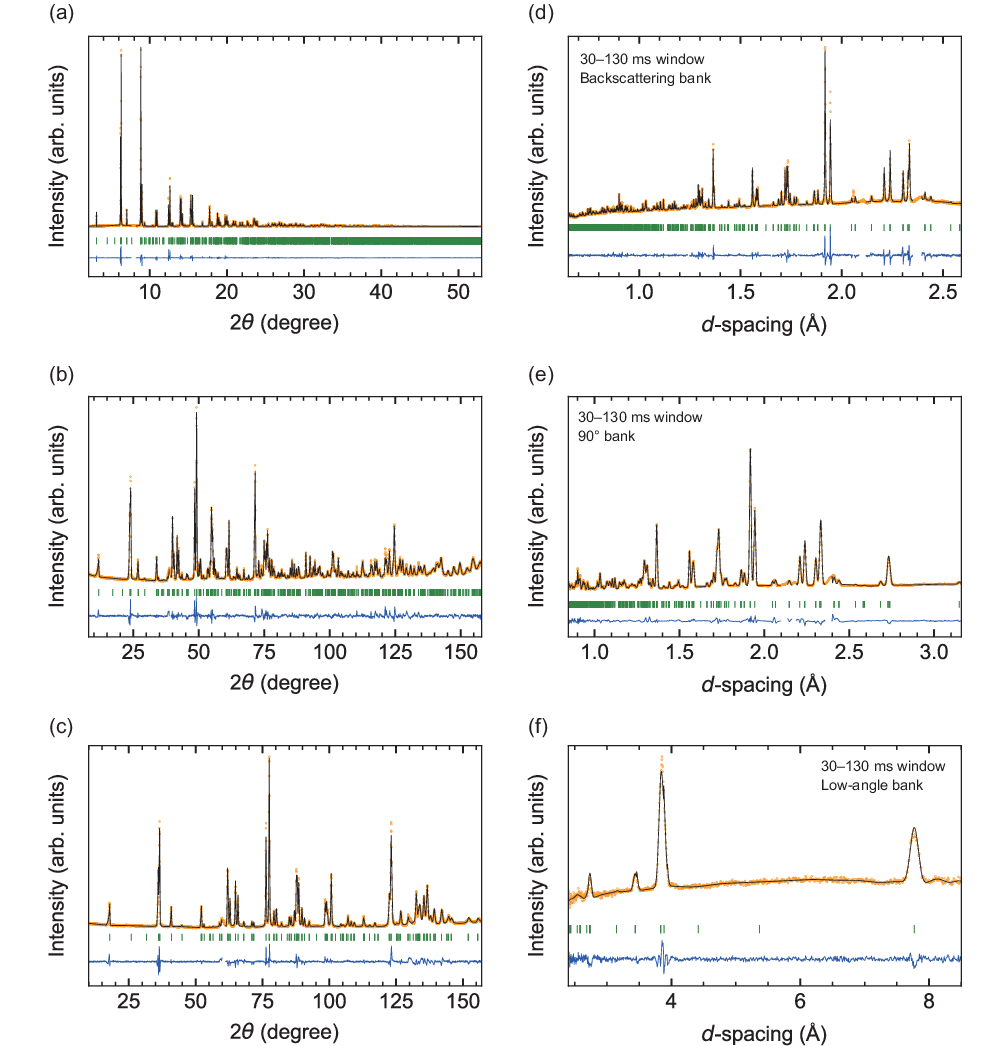}
   \caption{
   Rietveld refinement results for 300-K (a) SXRD ($\lambda$= 0.42112 \AA), (b, c) 2$\theta$-dispersive ND, and (d--f) time-of-flight ND data of Y$_{1/3}$TaO$_3$ with a $Pb2_1m$ structural model.
   Orange circles, black lines, and blue lines represent the observed, calculated, and diﬀerence profiles, respectively.
   The green ticks indicate the position of Bragg reflections.
   All six data sets were fitted simultaneously.
   Broad satellite peaks are excluded from the refinement.
   }
\label{fig:riet_300K}
\end{figure*}

The observed reflection conditions in the superlattice setting are 0$k$$l$: $k = 2n$ and 0$k$0: $k = 2n$, where $n$ is an integer,
leading to the extinction symbol of $Pb--$ and suggesting candidate space groups of $Pbm2$ (No.~28), $Pb2_1m$ (No.~26), and $Pbmm$ (No.~51). 
Among these,  $Pbmm$ and $Pb2_1m$ are compatible with defect-ordered perovskites structures as summarized in Ref.~\onlinecite{howard2004ACB}.
Since Y$_{1/3}$TaO$_3$ exhibits a clear SHG signal at room temperature (as discussed later), we can definitively exclude the centrosymmetric $Pbmm$ space group. 
Hence, we assign the room-temperature structure of this perovskite compound to the polar $Pb2_1m$ space group.
The Rietveld fitting results for SXRD and ND patterns with $Pb2_1m$ symmetry are plotted in Fig.~\ref{fig:riet_300K}. 
This structural model accurately captures the diffraction intensity particularly when split-atom model is applied to the equatorial oxygen sites.
The final refined structural parameters are listed in Table.~\ref{tb:strct_param_300K}.
Occupation of Y at the vacancy layers was excluded from the refinement, as a HAADF-STEM image (\red{Fig.~S4 in Supporting Information (SI)}) confirmed a complete layer-by-layer alternation of Y and vacancies.

\begin{table*}
   \caption{
   Crystallographic parameters of Y$_{1/3}$TaO$_3$ at 300~K obtained from joint Rietveld fitting with a $Pb2_1m$ model to the SXRD and ND data.
   }
   \begin{ruledtabular}
   \begin{tabular}{lcccccc}
    Atom & Site & $x$        & $y$       & $z$        & occ.      & $U_{\mathrm{iso}}$ (\AA$^2$)                \\ \hline
    Y    & 2$a$ & 0.7435(3)  & 0.0313(4) & 0          & 0.666     & 0.0109(4)                                   \\
    Ta   & 4$c$ & 0.2475(3)  & 0         & 0.74033(8) & 1         & 0.00726(9)                                  \\
    O1   & 2$a$ & 0.8181(3)  & 0.4805(7) & 0          & 1         & 0.0267(6)                                   \\
    O2   & 2$b$ & 0.6734(3)  & 0.5009(8) & 1/2        & 1         & 0.0092(4)                                   \\
    O3a  & 4$c$ & 0.480(3)   & 0.262(2)  & 0.8068(7)  & 0.65(2)   & 0.0023(2)                                   \\
    O3b  & 4$c$ & 0.545(4)   & 0.240(3)  & 0.773(3)   & 0.162(19) & = $U_{\mathrm{iso}}$(O3a)                   \\
    O3c  & 4$c$ & 0.425(4)   & 0.309(4)  & 0.8067(14) & 0.19(3)   & = $U_{\mathrm{iso}}$(O3a)                   \\
    O4a  & 4$c$ & 0.0611(5)  & 0.6186(6) & 0.7303(3)  & 0.639(5)  & = $U_{\mathrm{iso}}$(O3a)                   \\
    O4b  & 4$c$ & 0.9711(10) & 0.8082(8) & 0.7255(5)  & 0.361(5)  & = $U_{\mathrm{iso}}$(O3a)                  
    \end{tabular}
   \end{ruledtabular}
   Space group: $Pb2_1m$ (No.~26), $Z = 4$. Cell parameters: $a$ = 5.36946(3)~\AA, $b$ = 5.47603(3)~\AA, and $c$ = 7.77164(4)~\AA.
   $R_{\mathrm{wp}}$ = 6.93\%, GoF = 2.99.
\label{tb:strct_param_300K}
\end{table*}

We should remark here that the 300-K ND pattern of Y$_{1/3}$TaO$_3$ contains a weak and broad diffraction peak in a $d$-range of 2.35--2.45~\AA\ and
even weaker one at 1.27~\AA,
which cannot be indexed by the $Pb2_1m$ model or others with $Pbmm$ and $Pbm2$ symmetry (\red{Fig. S1}).
Our investigation described in \red{Section S1.B of SI} reveals that these weak peaks originates from $(3 + 2)$-dimensional incommensurate modulation
likely inherited from the incommensurate paraelectric structure observed at higher temperatures (see the next subsection). 
Although we list possible superspace groups and modulation vectors for the room-temperature incommensurate structure,
quantitative refinement is impractical because of the weak and broad feature of the satellite peaks,
probably indicating a short coherent length of the modulation.
Our $Pb2_1m$ model thus represents an average structure, where the modulated nature is imitated by the split-atom model, but remains reliable for discussing the structural features, given the good fitting to the basic reflections. 

\begin{figure}
   \includegraphics[width=8.6cm]{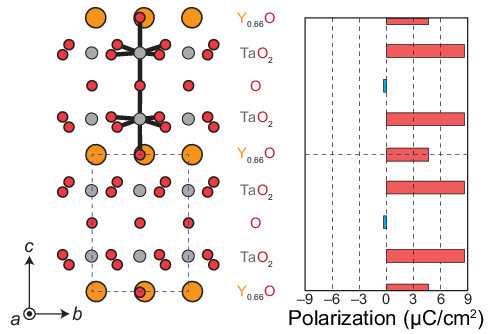}
   \caption{
   Atomic structure illustration of the [100]-projected $Pb2_1m$ phase and the layer-resolved local polarization along the $b$ axis calculated for the refined $Pb2_1m$ structure with a point-charge approximation. 
   The blue dashed square represents the unit cell.
   }
\label{fig:resolved}
\end{figure}

Electronic polarization of Y$_{1/3}$TaO$_3$ is calculated from the refined structural model with a point-charge approximation,
and the layer-resolved values of the electric dipoles are plotted in Fig.~\ref{fig:resolved}. 
As in Fig.~\ref{fig:structures_scheme}, the local polarization at the vacancy layer (O layer) is marginal due to the absence of A-site cation.
Importantly, the polarization profile is almost ferroelectric---rather than \textit{ferrielectric}---with significant contributions from TaO$_2$ and Y$_{0.66}$O layers.
The net macroscopic polarization is estimated to be 21.8~$\mu$C/cm$^{2}$,
larger than those of canonical layered hybrid improper ferroelectrics such as Ca$_3$Ti$_2$O$_7$ (11.8~$\mu$C/cm$^{2}$)~\cite{senn2015PRL} and Sr$_3$Zr$_2$O$_7$ (6.75~$\mu$C/cm$^{2}$)~\cite{yoshida2018AFM} 
as well as an improper multiferroic YMnO$_3$ ($\simeq$ 3~$\mu$C/cm$^{2}$ at 300~K)~\cite{gibbs2011PRBa}
This enriched polarization and its unique layer-resolved profile  corroborate our idea of utilizing vacancy ordering.
A similar strategy focusing on vacancy engineering can be applied to other (hybrid) improper ferroelectrics, including magnetoelectric multiferroics,
to enhance their electric polarization.

\begin{figure}
   \includegraphics[width=8.6cm]{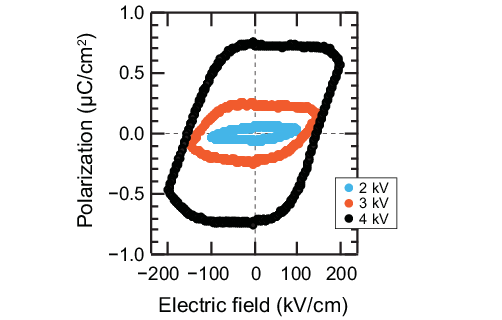}
   \caption{
   $P$--$E$ hysteresis loops measured for a polycrystalline Y$_{1/3}$TaO$_3$ pellet at room temperature.
   Electric field was applied at a frequency of 1~Hz. 
   }
\label{fig:PE}
\end{figure}

Compelling evidence for ferroelectricity can only be provided by polarization switching experiments.
Our $P$--$E$ measurements for dense pellets of Y$_{1/3}$TaO$_3$ yield the polarization curves displayed in Fig.~\ref{fig:PE}.
The measured curves unequivocally demonstrate polarization hysteresis as a function of the electric field,
confirming the ferroelectric character of Y$_{1/3}$TaO$_3$. 
The net electric polarization and coercive field are $\simeq$0.8~$\mu$C/cm$^{2}$ and $\simeq$150~kV/cm, respectively.
These values fall within the range reported for ceramic samples of hybrid improper ferroelectrics: 
0.9~$\mu$C/cm$^{2}$ and 110~kV/cm for Ca$_3$Ti$_2$O$_7$~\cite{hu2019JoM},
0.62~$\mu$C/cm$^{2}$ and 250~kV/cm for Sr$_3$Sn$_2$O$_7$~\cite{chen2020APL},
and
0.3~$\mu$C/cm$^{2}$ and 150~kV/cm for Sr$_3$Zr$_2$O$_7$~\cite{yoshida2018AFM}.
While the point-charge evaluation of the $Pb2_1m$ structure suggests a larger intrinsic polarization, 
we regard the experimental value as a lower bound for at least three reasons:
(i)~grain-boundary effect,
(ii)~domain-wall pinning,
and
(iii)~residual incommensurate modulation at room temperature.
To mitigate the first reason, further optimization of the synthesis and sintering process, similar to strategies successfully employed in other hybrid improper ferroelectrics~\cite{hu2021JoAaC},
is a promising route.
Approaches to address the other reasons and realize the intrinsic polarization potential of Y$_{1/3}$TaO$_3$ are discussed below after revealing the microscopic character of its paraelectric phase.

\subsection{$(3 + 2)$-dimensional incommensurate paraelectric structure}
\begin{figure}
   \includegraphics[width=8.6cm]{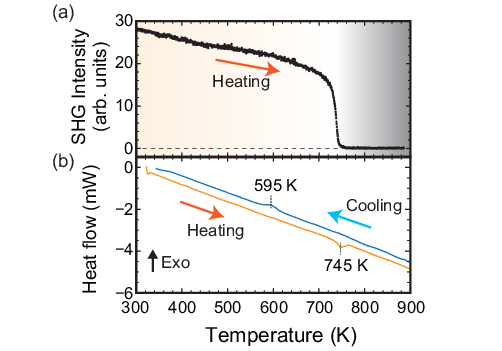}
   \caption{
   (a) Temperature dependence of SHG intensity recorded for Y$_{1/3}$TaO$_3$ on heating.
   (b) DSC curves on heating (orange) and natural cooling (blue). 
   }
\label{fig:SHG_DSC}
\end{figure}
Motivated to resolve how our chemistry-engineered balance between competing polar and rotational instabilities evolves with temperature,
we performed temperature-dependent SHG measurements to probe the structural evolution of Y$_{1/3}$TaO$_3$.
First, a finite SHG signal observed at room temperature conclusively rules out centrosymmetric $Pbmm$ structure as a possibility,
supporting our assignment of the room-temperature phase to the polar $Pb2_1m$ space group.
As shown in Fig.~\ref{fig:SHG_DSC}a, the SHG intensity decreases as temperature increases, becoming undetectable above 740~K, which implies a phase transformation to a centrosymmetric paraelectric phase.
This finding is consistent with the DSC curve (Fig.~\ref{fig:SHG_DSC}b), which reveals an endothermic peak at 745~K during heating.
Conversely, an exothermic peak was observed at 595~K during the cooling cycle.
The presence of latent heat and the significant thermal hysteresis between heating and cooling cycles suggest a first-order nature for this ferroelectric-to-paraelectric transition.

To assess the atomic structure of the paraelectric phase, we next inspect the temperature-variable SXRD and ND patterns. 
While the disappearance of superlattice reflections and peak splitting (indicative of the enlarged orthorhombic unit cell) above 800~K
suggests a return to the tetragonal aristotype structure with space group $P4/mmm$ (Fig.~\ref{fig:SLR}), 
we observed additional Bragg reflections in ND patterns that could not be indexed with the aristotype unit cell (Fig.~\ref{fig:SLR}c).
These reflections are also present with extremely low intensity in high-temperature SXRD patterns  (\red{Fig. S1}).
Notably, these weak reflections appear at different positions than expected for superlattice reflections corresponding to the
$\sqrt{2}\bm{a}_{\mathrm{pc}} \times \sqrt{2}\bm{b}_{\mathrm{pc}} \times 2\bm{c}_{\mathrm{pc}}$ cell metric
and have distinct $d$-spacing from the satellite observed at 300~K.
As shown in \red{Fig.~S1}, these additional reflections disappear on the subsequent cooling to 300~K,
indicating that they are not impurity peaks due to the decomposition and/or degradation of the sample. 
Commensurate superlattices, such as
$2\bm{a}_{\mathrm{pc}} \times 2\bm{b}_{\mathrm{pc}} \times 2\bm{c}_{\mathrm{pc}}$,
$2\sqrt{2}\bm{a}_{\mathrm{pc}} \times 2\sqrt{2}\bm{b}_{\mathrm{pc}} \times 2\bm{c}_{\mathrm{pc}}$, and
$2\sqrt{2}\bm{a}_{\mathrm{pc}} \times 2\sqrt{2}\bm{b}_{\mathrm{pc}} \times 4\bm{c}_{\mathrm{pc}}$,
do not offer reasonable Miller indices to these reflections.

\begin{figure}
   \includegraphics[width=8.6cm]{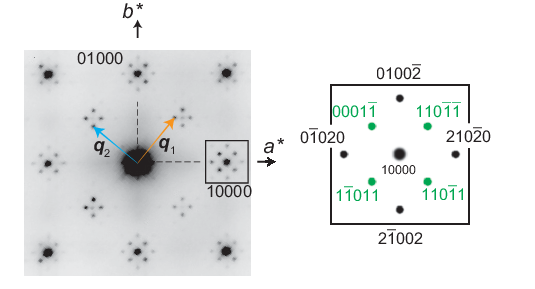}
   \caption{
   SAED pattern of Y$_{1/3}$TaO$_3$ along [001] zone axis at 900~K.
   Orange and blue arrows represent the modulation vectors, $\bm{q}_1$ and $\bm{q}_2$, respectively.
   An enlarged indexing scheme around the 10000 basic reflection with respect to the $(3 + 2)$-dimensional superstructure is shown on the right side. 
   }
\label{fig:satellites_SAED}
\end{figure}

Figure~\ref{fig:satellites_SAED} shows the SAED pattern along the [001] zone axis at 900~K
and demonstrates the presence of satellite reflections close to the basic spots, for example, 100 reflection highlighted by a black square.
All the SAED spots observed for Y$_{1/3}$TaO$_3$ at 900~K can be indexed by choosing the scattering vector $\bm{H}$ of
\begin{equation}
    \bm{H} = h \bm{a}^* + k \bm{b}^* + l \bm{c}^* + m \bm{q}_1 + n \bm{q}_2,
\end{equation}
where a set of \textit{hklmn} is Miller indices, $\bm{a}^*$, $\bm{b}^*$, and $\bm{c}^*$ are reciprocal lattice vectors of $P4/mmm$ lattice,
and $\bm{q}_1$ and $\bm{q}_2$ are modulation vectors represented by the equations below.
\begin{equation}
\bm{q}_1 = \alpha \bm{a}^* + \frac{1}{2} \bm{b}^*, 
\label{eq:q1}
\end{equation}
and
\begin{equation}
\bm{q}_2 = -\frac{1}{2} \bm{a}^* + \alpha \bm{b}^*,
\label{eq:q2}
\end{equation}
where $\alpha$ is an irrational number close to 0.414.
This scattering vector corresponds to $(3 + 2)$-dimensional modulation.

Based on a detailed analysis of SXRD, ND, and SAED patterns (see \red{section S1.A, SI}), we assign the paraelectric phase to a (3 + 2)-dimensional superspace group of 
$P4/mmm(\alpha,\frac{1}{2},0)0000(-\frac{1}{2},\alpha,0)0000$ (No.~123.2.64.16 in Stokes--Campbell--van Smaalen notation~\cite{stokes,stokes2011ACA,vansmaalen2013ACA}). 
The Rietveld refinement against the SXRD and ND data at 900~K results in the fitting displayed in \red{Fig.~S3},
providing a good statistic (Overall $R_{\mathrm{wp}}$ = 2.76\% and GoF = 2.30).
The obtained crystallographic parameters and the Fourier components of the modulation are listed in \red{Tables S2, S3, and S4}.

\begin{figure*}
   \includegraphics[width=17cm]{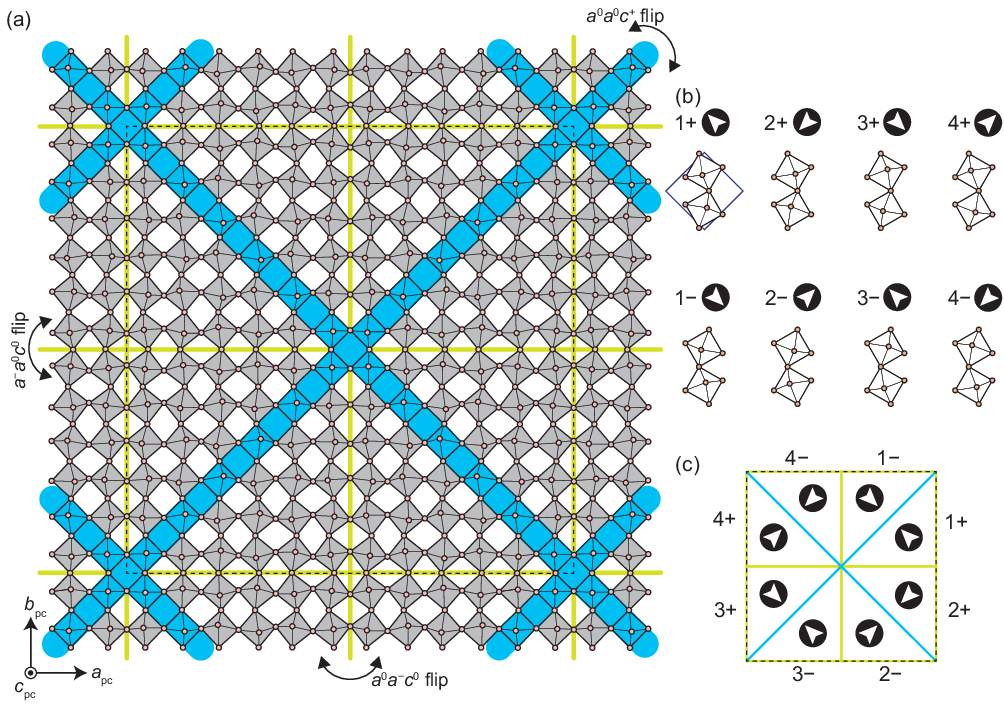}
   \caption{
   (a) Commensurately approximated structure of the paraelectric phase of Y$_{1/3}$TaO$_3$. Y atoms with small modulation are omitted for clarity.
   Black dashed, green, and blue lines represent the $12\bm{a}_{\mathrm{pc}} \times 12\bm{b}_{\mathrm{pc}} \times 2\bm{c}_{\mathrm{pc}}$ unit cell,
   twin boundaries (also known as ferroelastic tilting (FA$_{\mathrm{t}}$) domain walls), and ferroelectric rotation (FE$_{\mathrm{r}}$) domain walls, respectively. 
   (b) Labeling scheme for the eight degenerate states with $a^-a^-c^+$-type distortion.
   The blue square represents the $\sqrt{2}\bm{a}_{\mathrm{pc}} \times \sqrt{2}\bm{b}_{\mathrm{pc}} \times 2\bm{c}_{\mathrm{pc}}$ supercell seen along the $c$ direction.
   The white arrows indicate the direction of macroscopic polarization for each variant.
   (c) A proposed domain configuration for the commensurate approximation.
   We follow Ref.~\onlinecite{huang2016NC} to classify and name the domains and degenerate states. 
   }
\label{fig:approx_str}
\end{figure*}

Incommensurately modulated structures, which do not repeat periodically in physical space, pose challenges for direct visualization.
However, a commensurate approximation offers a feasible route to perceive their intricate atomic arrangements.
For the 900-K structure, the modulation vectors, $\bm{q}_1$ and $\bm{q}_2$, have the form of Eqs.~(\ref{eq:q1}) and (\ref{eq:q2}),
where the $\alpha$ value was determined to be 0.41291(7) (\red{section~S1.A, SI}).
Since this $\alpha$ is close to a rational number $\frac{5}{12}$ = $0.4166 \cdots $,
we construct an approximated structure of the paraelectric phase using the following $q$-vectors:
\begin{equation}
   \bm{q}_1^{\mathrm{C}} = \frac{5}{12} \bm{a}^* + \frac{1}{2} \bm{b}^*,
\end{equation}
and
\begin{equation}
   \bm{q}_2^{\mathrm{C}} = - \frac{1}{2} \bm{a}^* + \frac{5}{12} \bm{b}^*.
\end{equation}
Figure~\ref{fig:approx_str} illustrates this commensurate approximation within a $12\bm{a}_{\mathrm{pc}} \times 12\bm{b}_{\mathrm{pc}} \times 2\bm{c}_{\mathrm{pc}}$ supercell,
clearly revealing the displacive modulation of the oxide ions. 
Within this structure, the TaO$_6$ octahedra exhibit rotation about the $\bm{a}_{\mathrm{pc}}$, $\bm{b}_{\mathrm{pc}}$, and $\bm{c}_{\mathrm{pc}}$ axes
with the rotation amplitude varying from one octahedron to another.
For instance, the atomic arrangement at the corner of the $12\bm{a}_{\mathrm{pc}} \times 12\bm{b}_{\mathrm{pc}} \times 2\bm{c}_{\mathrm{pc}}$ cell resembles
the ideal $P4/mmm$ structure without octahedral rotations.
Conversely, other regions, particularly outside the immediate vicinity of the domain boundaries,
exhibit characteristics similar to the ferroelectric $Pb2_1m$ structure accompanied by $a^-a^-c^+$-type octahedral rotations.

The modulated paraelectric structure is characterized by distinct domain walls.
At the ferroelastic tilting domain walls (green lines in Fig.~\ref{fig:approx_str}a),
either $a^-a^0c^0$ or $a^0a^-c^0$ tilt component is flipped.
Similarly, the $a^0a^0c^+$-rotation reverses across the ferroelectric rotation domain walls (blue lines in Fig.~\ref{fig:approx_str}a).
The tilt and rotation amplitudes are consequently zero on average, resulting in an overall average space group of $P4/mmm$.
Complex domain-wall topologies have also been observed in ferroelectric states of hybrid improper ferroelectrics,
and their analogy will be discussed next.

\subsection{Analogy to ferroelectric domain topology}
Hybrid improper ferroelectrics host multiple order parameters whose higher-order coupling produces a characteristic $\mathrm{Z}_4 \times \mathrm{Z}_2$ domain topology,
as established in (Ca,Sr)$_3$(Ti,Mn)$_2$O$_7$~\cite{oh2015NM,huang2016NC}.
The Z$_4$ part originates from four $a^-a^-c^0$ octahedral-tilt variants, whereas the Z$_2$ reflects two senses of the in-phase $a^0a^0c^+$ rotation (i.e., clockwise or counter-clockwise),
combined to yield eight degenerate variants.
Within this topology, five types of domain walls are recognized: 
ferroelastic tilting (FA$_{\mathrm{t}}$),
ferroelastic tilting+rotation (FA$_{\mathrm{tr}}$),
ferroelectric tilting (FE$_{\mathrm{t}}$),
and
ferroelectric rotation (FE$_{\mathrm{r}}$) domain walls
as well as
antiphase boundaries (APBs). 
Ferroelastic walls correspond to 90$^{\circ}$ reorientations of the polarization, while the polarization flips 180$^{\circ}$ in the adjacent domain separated by the ferroelectric domain walls. 
We adopt this taxonomy below when analyzing the modulated texture of Y$_{1/3}$TaO$_3$ given its similarity to the hybrid improper ferroelectric states, i.e., local $a^-a^-c^+$-type distortion.

The paraelectric phase can be parsed locally into eight variants with labels 1$\pm$, 2$\pm$, 3$\pm$, or 4$\pm$ (Fig.\ref{fig:approx_str}b) following Ref.~\onlinecite{huang2016NC}.
Figure~\ref{fig:approx_str}c shows a domain configuration within the $12\bm{a}_{\mathrm{pc}} \times 12\bm{b}_{\mathrm{pc}} \times 2\bm{c}_{\mathrm{pc}}$ unit,
in which all eight variants are identified.
Not all five possible domain walls, however, are realized: only FA$_{\mathrm{t}}$ and FE$_{\mathrm{r}}$ domain walls exist. 
This is consistent with the reported energy hierarchy for A$_3$B$_2$O$_7$-type hybrid improper ferroelectrics~\cite{huang2016NC},
FA$_{\mathrm{t}}$ $\leq$ FE$_{\mathrm{r}}$ $\leq$ FE$_{\mathrm{t}}$ $\leq$ FA$_{\mathrm{tr}}$ $\leq$ APB,
so that the two lowest-energy domain walls dominate the incommensurate paraelectric phase. 
APBs, namely, interfaces between distinct domains sharing the same polarization direction such as 1$+$ and 3$-$ pair, are highest in energy and absent in the domain configuration of Fig.~\ref{fig:approx_str}.

Upon cooling from the incommensurate state, Y$_{1/3}$TaO$_3$ transforms to the ferroelectric $Pb2_1m$,
with polar domains nucleating within the pre-existing modulated matrix.
In conventional hybrid improper ferroelectrics, ferroelastic domains are micron scale~\cite{oh2015NM},
and the ferroelectric domain walls and APBs that develop inside them typically span hundreds of nanometers to micrometers.
In Y$_{1/3}$TaO$_3$, however, the ferroelastic domains are much smaller (green-outlined region in Fig.~\ref{fig:approx_str}, $\simeq$ $6\bm{a}_{\mathrm{pc}} \times 6\bm{b}_{\mathrm{pc}}$ $\simeq$ 2.3 $\times$ 2.3~nm$^2$),
so 180$^{\circ}$ ferroelectric walls and APBs that nucleate therein inherit this short length scale and remain fragmented.
Because APBs serve as a key nucleation center for polarization reversal in hybrid improper systems~\cite{huang2016NC},
such a shortened APB network impedes the switching and reduces the effectively switched volume fraction.
This behavior corresponds to (ii) domain-wall pinning, and we attribute the modest observed polarization to this constraint imprinted from the pre-existing incommensurate phase.

\begin{figure}
   \includegraphics[width=8.6cm]{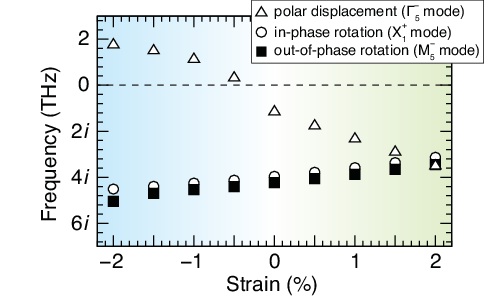}
   \caption{
   Calculated phonon-mode frequencies of polar displacement and rotational modes as a function of epitaxial strain.
   The calculations were performed for the model Ca$_{1/2}$TaO$_3$ system.
   }
\label{fig:strain}
\end{figure}

The residual incommensurate signatures at room temperature (\red{section S1.B, SI}) can be understood as a quenched remnant of the high-temperature modulation,
upon which polar domains are superimposed.
Therefore, eliminating the modulation from the paraelectric phase simultaneously removes (ii) domain-wall pinning and (iii) residual room-temperature modulation,
thereby recovering the intrinsically enhanced polarization of Y$_{1/3}$TaO$_3$.
A possible route to achieve this is to further unbalance the polar-vs-rotational competition toward the rotational side.
We here propose epitaxial strain as a practical knob.
As shown by our DFT calculations under biaxial strain (Fig.~\ref{fig:strain}), compressive strain enhances the rotational phonon instabilities while the polar mode hardens to real frequency,
removing the polar--rotational competition.
The required strain magnitude lies within experimentally accessible ranges ($\simeq$ 1--2\%)~\cite{choi2004S,lee2010N}.
By contrast, tensile strain shows the opposite trend, softening the polar mode and stiffening the rotations, which restores the competition and thus stabilizes the incommensurate modulation even at lower temperatures.
Fabricating Y$_{1/3}$TaO$_3$ in thin-film form is an interesting future direction to validate this route and further demonstrate the potential of this material.
Furthermore, compensated incommensurate phases, such as those observed in A$_{1-x}$BO$_3$~\cite{beqiri2019CC,zhou2008CM,labeauu1982ACSA}, are also seen in double-perovskite systems with polar-distortion active cations at the B site~\cite{abakumov2013CM,garcia-martin1999JoSSC,garcia-martin2008JACS,zuo2019IC}.
Since these architectures are intrinsically ready to show hybrid improper ferroelectricity once a commensurate $a^-a^-c^+$-type distortion is realized~\cite{benedek2011PRL,mulder2013AFM,young2013CM},
our chemistry- and strain-based engineering to unbalance the mode competition is also a promising strategy for these systems to lock into a commensurate polar phase,
thereby enriching the pool of hybrid improper ferroelectrics.

\section{Conclusion}
In this work, we report the first example of a defect perovskite with switchable ferroelectricity:
Y$_{1/3}$TaO$_3$, whose polar average $Pb2_1m$ structure is established by the combination of vacancy ordering and octahedral rotations.
While the polar instability owing to B-site cations, such as Ti$^{4+}$ and Nb$^{5+}$, often drives the system into an undesirable centrosymmetric incommensurate phase
when competing with rotational instabilities,
our design strategy mitigates this by using Ta$^{5+}$, which is less prone to polar displacement while maintaining the rotational instability.
The alternating arrangement of Y$^{3+}$ and vacancies creates a net ferroelectric dipole ordering, giving rise to a polarization larger than that in conventional \textit{ferrielectric}-like improper systems.
Although the measured polarization value ($\simeq$0.8~$\mu$C/cm$^2$) is comparable to known hybrid improper ferroelectrics,
our detailed structural analysis of the paraelectric phase and DFT calculations suggest that an intrinsically enhanced polarization ($\simeq$21.8~$\mu$C/cm$^2$) could be realized
by removing the modulation in the paraelectric phase via moderate compressive strain (1--2\%).
We anticipate that this defect-ordering approach provides a route to amplify polarization in a wide range of improper ferroelectrics,
including magnetoelectric multiferroics.
Furthermore, the concept of resolving competing instabilities, validated here in the A$_{1-x}$BO$_3$ family, should be applicable to AA$^{\prime}$B$_2$O$_6$ and AA$^{\prime}$BB'O$_6$ double perovskites,
where incommensurate centrosymmetric ground states are also observed frequently.
This opens a pathway to stabilizing hybrid improper ferroelectric states and to discovering new ferroelectric materials.

\section{Supporting Information}
The Supporting Information is available free of charge at URL.
\begin{itemize}
   \item Details of structural analysis of 300-K and 900-K phases, including additional SXRD and ND data, superspace formalism, Rietveld refinement profiles,
         crystallographic parameters, and SAED (PDF).
\end{itemize}

\section{acknowledgments}
The authors thank Dr.~Clemens Ritter and Dr.~Alexandra S. Gibbs for their help with the ND measurements.
This work was supported by Japan Science and Technology Agency (JST)
as part of Adopting Sustainable Partnerships for Innovative Research Ecosystem (ASPIRE), Grant Number JPMJAP2408,
and JSPS KAKENHI, Grant Numbers JP24K21688 and JP25K01500.
SHG measurements were supported by the US Department of Energy, Basic Energy Sciences, Grant Number DE-SC0012375.
$2\theta$-dispersive ND experiments on D2B at the Institute Laue Langevin were performed under proposal No.~5-23-714.
Time-of-flight ND experiments on HRPD at the ISIS Pulsed Neutron and Muon Source were supported by beamtime allocation from STFC (RB1820148).
%

%
\end{document}